\documentclass[10pt,a4paper]{article}
\usepackage[latin1]{inputenc}
\usepackage{amsmath}
\usepackage{authblk}
\usepackage{amsfonts}
\usepackage{amssymb}
\usepackage{graphicx}
 \usepackage[usenames,dvipsnames]{pstricks}
 \usepackage{epsfig}
 \usepackage{pst-grad} % For gradients
 \usepackage{pst-plot} % For axes
\author[1]{O. Raz}
\author[1]{Y. Suba\c{s}\i} 
\author[1,2]{C. Jarzynski}
\affil[1]{Department of Chemistry and Biochemistry ,
	University of Maryland, College Park, MD 20742, U.S.A.}
\affil[2]{Institute for Physical Science and Technology,
	University of Maryland, College Park, MD 20742, U.S.A.}
\title{Mimicking  Nonequilibrium Steady States  with Stochastic Pumps}
\begin{document}
\maketitle
\begin{abstract}

We establish a correspondence between two very general paradigms for systems that persist away from thermal equilibrium. In the first paradigm, a nonequilibrium steady state (NESS) is maintained by applying fixed thermodynamic forces that break detailed balance.
In the second paradigm, known as a stochastic pump (SP), a time-periodic state is maintained by the periodic variation of a system's external parameters.
In both cases, currents are generated and entropy is produced.  Restricting ourselves to discrete-state systems, we establish a mapping between these scenarios. Given a NESS characterized by a particular set of stationary probabilities, currents and entropy production rates, we show how to construct a SP with exactly the same (time-averaged) values.  The mapping works in the opposite direction as well. 
These results establish an equivalence between the two paradigms, by showing that stochastic pumps are able to mimic the behavior of nonequilibrium steady states, and vice-versa.
\end{abstract}
\section{Introduction and Motivation}

While there is currently no single theory that unifies all nonequilibrium phenomena, a number of useful paradigms of nonequilibrium behavior have emerged.
These include: small perturbations near equilibrium;
systems driven away from an initial state of equilibrium;
spontaneous relaxation towards equilibrium;
non-equilibrium steady states generated by fixed thermodynamic forces;
and stochastic pumps driven by the time-periodic variation of external parameters.
Theoretical frameworks developed within each paradigm -- for instance, linear response theory to describe near-equilibrium perturbations -- have contributed to a broader understanding of nonequilibrium processes.

In this work we focus on two of these paradigms: {\it non-equilibrium steady states} and {\it stochastic pumps}.
These share certain features, notably the persistence of non-vanishing currents and entropy production, which invite a comparison between the two.
As elaborated below, we will devise a mapping from one paradigm to the other: given a system in a nonequilibrium steady-state characterized by certain occupation probabilities, currents and entropy production rates, we will show how to construct a stochastic pump that exhibits the same (time-averaged) properties. The inverse direction, namely the construction of a nonequilibrium steady state with the same properties as a given, time-averaged stochastic pump, will also be discussed.

In the nonequilibrium steady-state (NESS) paradigm, a system driven by fixed thermodynamic forces -- such as temperature gradients or chemical potential differences --
reaches a steady state in which its statistical properties are stationary with time.
Unlike an equilibrium state, a nonequilibrium steady state exhibits non-vanishing currents, reflecting the violation of detailed balance.
In order to maintain such a state, a thermodynamic cost must be paid.
This cost is measured by the continual depletion of a thermodynamic resource, such as a chemical fuel, resulting in the production of entropy in the system's thermal surroundings.

Biomolecular motors illustrate the NESS paradigm~\cite{Bio_Molecular_motor_review,Kolomeisky2007}: a reaction such as $ATP$ hydrolysis ($ATP \rightarrow ADP + P_i$) produces entropy in the surrounding solution, and the chemical potential difference between reactants and products provides the thermodynamic force.
The ``current'' in this situation corresponds to the mechanical motion produced by the motor, for instance the systematic displacement of kinesin motor toward the positive end of a microtubule filament.
For a recent review of the stochastic theory of nonequilibrium steady states, as applied to biochemical processes, see Ref.~\cite{NESS_review_1}.

In the stochastic pump paradigm, a system is driven by the time-periodic variation of external parameters in the presence of a thermal reservoir.
Typically, it is assumed that the dynamics satisfy detailed balance at every instant in time -- in other words, if the parameters were suddenly frozen at their instantaneous values, the system would relax to an equilibrium state.
Under suitable conditions, a periodically driven system reaches a time-periodic state with non-vanishing time-averaged currents.
These currents are effectively ``pumped'' by the periodic variation of the parameters, and the cost associated with pumping these currents is the work invested in driving the parameters.
Ultimately, the energy provided by this work is dissipated into the thermal reservoir, resulting in the production of entropy.

The study of stochastic pumps has been stimulated by experiments on artificial molecular machines {\cite{Artificial_molecular_motors,Experimental_Engine}, which are manipulated by the variation of external parameters to achieve some desired behavior. For instance, in experiments on catenanes -- mechanically interlocked ring-like molecules -- the aim was to produce unidirectional rotation of one ring around the other~\cite{Leigh2003}.
Theoretical investigations of  SP have focused on slowly driven \cite{Adiabatic_Molecular_motors_2007,sinitsyn2007berry} as well as weakly driven \cite{Perturbative_Pumps} pumps, ``no-pumping'' theorems \cite{mandal2011No_pump_graphical,rahavNo_pump_PRL,chernyak_no_pump_2008,No_pump_maes_2010,No-pump_mandal2014unification}, the role of interactions \cite{Rahav_No_Pump_with_interactions} and fluctuations \cite{ren2011duality}, and the ability to extract work from stochastic pumps \cite{rahav2011extracting}.
%``produce controlled motion at the molecular level'' (Leigh 2003 abstract)

Underlying both the experimental work on artificial molecular machines and the theoretical work on stochastic pumps is the broad goal of understanding how to achieve controlled motion at the molecular level, where thermal fluctuations are large.
The focus on time-dependent driving is motivated in part by the difficulty of synthesizing artificial molecular systems that, like biological molecular motors, takes advantage of chemical potential differences to drive steady currents.
It is often simpler to manipulate the system by varying external parameters such as temperature and the surrounding chemical environment.

In both nonequilbrium steady states and stochastic pumps, the generated currents can be viewed as desired outcomes, and entropy production as the cost of achieving them.
In this perspective, fixed thermodynamic forces (NESS) and time-periodic external driving (SP) represent the tools at our disposal.
It is then interesting to compare these two sets of tools with respect to the degree of control that can be achieved.
In particular, in this paper we investigate whether time-periodic driving can always achieve the same time-averaged outcome (i.e. identical currents) as fixed thermodynamic forces, and at the same cost (identical time-averaged entropy production).
In other words, can a stochastic pump ``mimic'' an arbitrary nonequilibrium steady state?

More precisely, we begin by considering a generic Markov process of random transitions among $n$ states of a system.
The transition rates are fixed in time and do not satisfy detailed balance, hence the dynamics lead to a NESS with nonvanishing currents and entropy production.
We then show how to prescribe a stochastic pump that has, in the limit of many cycles, the same time averaged probabilities, currents and entropy production rates as the NESS.
This prescription is constructive though not unique.
Surprisingly, the construction does not require the solution of any differential equations, only linear, algebraic equations.
By contrast, a mapping in the opposite direction (from SP to NESS) requires that we first determine the periodic state of the system, which involves solving a set of coupled ordinary differential equations with time-periodic parameters.
Typically this can only be done numerically.

The rest of the manuscript is organized as follows: In section \ref{Sec:Setup} we formulate the problem and review some useful known results. We then consider in section \ref{Sec:FromSP2NESS} the simpler direction, namely mapping a stochastic pump into a NESS. In section  \ref{Sec:KeyIdea} we present two types of transformations for detailed balance matrices which play a key role in our construction of a SP that mimics NESS, and the ``no current-loops'' property that sets a constraint on pumping with time-dependent detailed balance matrix. The construction of a pumping protocol is described in section \ref{Sec:Construction}. We finish with few concluding remarks in section \ref{Sec:Conclusion}.
                
\section{Setup and Background} \label{Sec:Setup}
\subsection{Definitions} \label{Subsec:Defs}
We limit ourselves to an ergodic, continuous-time Markovian system with $n$ states.
The evolution of the system consists of random, Poissonian transitions among these states, with transition rates that are governed either by a time-independent rate matrix $\mathcal{R}$ (when analyzing nonequilibrium steady states) or a time-periodic rate matrix $\mathcal{W}(t)$ (for stochastic pumps), as described in more detail below.
It is convenient to picture the system in terms of a ``connectivity graph'' with $n$ nodes and a finite number of edges connecting given pairs of nodes. An edge between states $i$ and $j$ implies that the system can make transitions between these states (see Fig. \ref{fig:Figure_General_Graph}).
We will use $\vec p(t)$ to denote the vector of the probabilities whose $i$'th component $p_i(t)$ is the probability for the system to be in state $i$ at time $t$.
\begin{figure}
\centering
\includegraphics[width=0.5\linewidth]{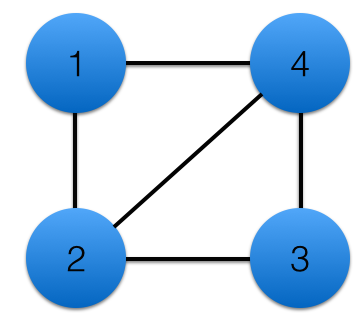}
\caption{A four-state system described by a graph. Each node represents a state of the system. The edges represent non-vanishing transition rates between states. In this example, direct transitions between states 1 and 3 are not allowed.}
\label{fig:Figure_General_Graph}
\end{figure}

In the NESS scenario, the evolution of the system obeys the master equation
\begin{eqnarray}\label{eq:Master}
\partial_t \vec p = \mathcal{R}\vec p.
\end{eqnarray}   
For $i\neq j$, the matrix element $\mathcal{R}_{ij}$ is the probability per unit time for a system in state $j$ to make a transition to state $i$. The diagonal elements of $\mathcal{R}$ are negative, and are determined by conservation of probability: $\sum_i \mathcal{R}_{ij} =0$. For the system to be ergodic, we demand that (i) if $\mathcal{R}_{ij}\neq0$ then also $\mathcal{R}_{ji}\neq 0$, and (ii) the graph associated with $\mathcal{R}$ is connected.
That is, for any pair of nodes (states) $i$ and $j$, there exists a path from $i$ to $j$, possibly through a sequence of intermediate nodes, along the edges of the graph.
Under these conditions, Eq.(\ref{eq:Master}) has a unique steady state solution, which we will denote by $\vec p^{\,ss}$, and any solution of Eq.(\ref{eq:Master}) converges to this steady state in the long-time limit~\cite{schnakenberg1976network}.

In addition to the steady state probabilities $\vec p^{\,ss}$, we will be interested in the steady state currents, defined by 
\begin{eqnarray}\label{Eq:CurrentsDef}
\mathcal{J}_{ij}^{ss} = \mathcal{R}_{ij}p^{ss}_j - \mathcal{R}_{ji}p^{ss}_i,
\end{eqnarray}
and the entropy production rates associated with these currents~\cite{schnakenberg1976network}:
\begin{eqnarray}\label{eq:SigmaDef}
\mathcal{\sigma}_{ij}^{ss} = \mathcal{J}_{ij}^{ss}\log\frac{\mathcal{R}_{ij}p_j^{ss}}{\mathcal{R}_{ji}p_i^{ss}}.
\end{eqnarray}
We will assume that some of these currents, and therefore the corresponding entropy production rates, are non-vanishing.
In other words we assume the dynamics violate detailed balance, hence $\vec p^{\,ss}$ is a genuinely {\it nonequilibrium} steady state.

In the stochastic pump scenario, the system obeys a master equation with a time-periodic rate matrix,
\begin{eqnarray}\label{eq:PerMaster}
\partial_t \vec p = \mathcal{W}(t)\vec p,
\end{eqnarray}
where
\begin{equation}
\label{eq:periodic}
\mathcal{W}(t) = \mathcal{W}(t+T)
\end{equation}
for some finite period $T$.
If we momentarily treat $t$ in Eq.(\ref{eq:PerMaster}) as a parameter of the rate matrix (rather than as the time variable), then for any fixed value of this parameter we will assume the rate matrix $\mathcal{W}(t)$ has a unique stationary solution $\vec \pi(t)$, and we further assume that
\begin{eqnarray}\label{Eq:DetailBalance}
\mathcal{W}_{ij} \pi_j  - \mathcal{W}_{ji} \pi_i  = 0\;\;\forall \, i,j.
\end{eqnarray}
In other words, the dynamics generated by $\mathcal{W}(t)$ (for fixed $t$) satisfy detailed balance.
We will refer to $\vec \pi(t)$ as the {\it equilibrium state} of $\mathcal{W}(t)$.
This is the state to which the system would relax if all the transition rates $\mathcal{W}_{ij}$ were ``frozen'' in time.

Let us now return to thinking of $t$ as time.
For any solution of Eq.(\ref{eq:PerMaster}) the quantities
\begin{eqnarray}\label{Eq:PerCurrentsDef}
\mathcal{J}_{ij}(t) &=& \mathcal{W}_{ij} p_j  - \mathcal{W}_{ji} p_i  \\
\label{eq:PerSigmaDef}
\mathcal{\sigma}_{ij}(t) &=& \mathcal{J}_{ij} \log\frac{\mathcal{W}_{ij} p_j }{\mathcal{W}_{ji} p_i }
\end{eqnarray}
(suppressing the argument $t$ on the right side) represent instantaneous currents and entropy production rates.
Under Eqs.(\ref{eq:PerMaster},\ref{eq:periodic}) the statistical state of the system evolves asymptotically to a unique time periodic state,
\begin{equation}
\vec p^{\,ps}(t) = \vec p^{\,ps}(t+T),
\end{equation}
with currents and entropy production rates
\begin{eqnarray}
\label{eq:Jps}
\mathcal{J}_{ij}^{ps}(t) &=& \mathcal{W}_{ij} p_j^{ps} - \mathcal{W}_{ji}p_i^{ps} \\
\label{eq:sigmaps}
\mathcal{\sigma}_{ij}^{ps}(t) &=& \mathcal{J}_{ij}^{ps}\log\frac{\mathcal{W}_{ij} p_j^{ps}}{\mathcal{W}_{ji} p_i^{ps}}.
\end{eqnarray}
These are analogous to the quantities appearing in Eqs.(\ref{Eq:CurrentsDef},\ref{eq:SigmaDef}), only $\mathcal{J}_{ij}^{ps}(t)$ and $\mathcal{\sigma}_{ij}^{ps}(t)$ are periodic with time, whereas $\mathcal{J}_{ij}^{ss}$ and $\mathcal{\sigma}_{ij}^{ss}$ do not vary with time.

Throughout this paper we will be interested in the asymptotic properties of the system, therefore we will consider only the steady state $\vec p^{\,ss}$ and the periodic state $\vec p^{\,ps}(t)$, and not the process of relaxation to either of these states.

In order to compare the NESS and stochastic pump (SP) scenarios, let us define the time averaged quantities in the periodic state of the SP:
\begin{equation}\label{Eq:TimeVaerageProb}
\begin{split}
\overline{p_{i}^{ps}} = \frac{1}{T}\int_0^Tp_{i}^{ps}(t)dt \qquad &,\qquad
\overline{\mathcal{J}_{ij}^{ps}}  =\frac{1}{T}\int_0^T\mathcal{J}_{ij}^{ps}(t)dt \\
\overline{\sigma_{ij}^{ps}} = \frac{1}{T} &\int_0^T\sigma_{ij}^{ps}(t)dt.
\end{split}
\end{equation}
The problem that we wish to study can now be formulated as follows.

\paragraph{Problem Formulation:}
Given a time-independent rate matrix $\mathcal{R}$ corresponding to steady state quantities $ \vec p^{\,ss}$, $\mathcal{J}^{ss}$ and $\sigma^{ss}$, we want to construct a time periodic detailed balance rate matrix $\mathcal{W}(t)$ whose periodic state is described by the same quantities, after averaging over time:
\begin{equation}\label{eq:goal}
\overline{p_{i}^{ps}} =p^{ss}_i \quad,\quad \overline{\mathcal{J}_{ij}^{ps}} = \mathcal{J}^{ss}_{ij} \quad,\quad \overline{\sigma^{ps}_{ij}} = \sigma^{ss}_{ij}.
\end{equation}
We denote the above problem as the ``forward'' problem. We will also be interested in the ``inverse'' problem: given a time-dependent detailed balance rate matrix $\mathcal{W}(t)$ corresponding to the time-averaged quantities $\overline{p_{i}^{ps}}$, $\overline{\mathcal{J}_{ij}^{ps}}$ and $\overline{\sigma^{ps}_{ij}}$, we want to construct a time-independent rate matrix $\mathcal{R}$ such that Eq.(\ref{eq:goal}) holds. 
As we discuss in more detail below, the solution of the inverse problem follows directly from known results, therefore we will focus mainly on the forward problem in this paper.

When $\mathcal{R}$ and $\mathcal{W}(t)$ give rise to dynamics that satisfy Eq.(\ref{eq:goal}), we will say that the stochastic pump ``mimics'' the nonequilibrium steady state, and vice-versa.

\subsection{Two Useful Decompositions}

The two well known decompositions described below, the first -- an algebraic decomposition of rate matrices, and the second -- a topological decomposition of the connectivity graph, will be extensively used in what follows. 

\subsubsection{Rate matrix decomposition}
The following (unique) decomposition of any rate matrix $\mathcal{R}$, obtained by Zia and Schmittmann~\cite{Schmittmann_zia_2007}, will prove to be useful:
\begin{eqnarray}\label{Eq:R_decomposition}
\mathcal{R} = \left(\mathcal {S} + \frac{1}{2}\mathcal{J}^{ss}\right)\cdot\mathcal{P}^{-1}.
\end{eqnarray} 
Here the multiplication is ordinary matrix multiplication, and $\mathcal {S}$ is a symmetric matrix whose elements in each column add up to zero, with negative entries only on the diagonal.
$\mathcal{J}^{ss}$ is the anti-symmetric current matrix defined in Eq.(\ref{Eq:CurrentsDef})  and  $\mathcal{P} = {\rm diag}(\vec p^{\,ss})$ is the diagonal matrix with elements $\mathcal{P}_{ii} = p^{ss}_i$.

An immediate corollary of Eq.~(\ref{Eq:R_decomposition}) is the following statement: if a rate matrix $\mathcal{R}$ is the product of a symmetric rate matrix $\mathcal {S}$ and a diagonal matrix $\mathcal {P}$ (with positive diagonal entries summing to unity), then $\mathcal{R}$ satisfies detailed balance, i.e.\ there are no currents in the stationary state.

\subsubsection{Cycle Decomposition}
The currents that characterize a NESS are in general not independent of one another, as they must satisfy constraints arising from the conservation of probability.
These constraints embody Kirchoff's law of currents.
The \emph{cycle decomposition} method provides a convenient tool to account for these constraints~\cite{schnakenberg1976network}.
Briefly, in a connected network with $N$ nodes and $E$ edges, the conservation of probability imposes $N-1$ constraints among the $E$ currents (one current per edge).
It is then convenient to identify $C=E-N+1$ {\it fundamental currents}, using the following procedure.
First, we build a connectivity graph for the system, as described above and illustrated in Fig.(\ref{fig:Figure_examp_system_NESS}) for the case of four nodes and six edges (hence $C=3$).
We then construct a maximal {\it spanning tree}, by removing $C$ edges without breaking the connectivity of the graph; this tree, illustrated by the red dash-dotted lines in Fig.(\ref{fig:Figure_examp_system_NESS}), has no cycles.
In the context of the original graph, the $C$ edges that are removed to form the spanning tree are called {\it fundamental edges}.  The currents along these edges
are the \emph{fundamental currents} (the black arrows in the figure), and the currents along the edges of the tree are the \emph{spanning tree currents} (the red arrows in the figure). 

For NESS, the steady state currents along the $C$ fundamental edges can take on any values, independently of one another.  However, once these fundamental currents are set,
the spanning tree currents are then uniquely determined by conservation of probability: the total sum of incoming and outgoing currents at each state must vanishes in the steady state. Therefore, the number of degrees of freedom in the matrix $\mathcal{J}^{ss}$ is not $(n^2-n)/2$ as for an arbitrary anti-symmetric matrix, but is determined by the graph topology. 

Unlike NESS, for stochastic pumps the fundamental currents  at each moment do not fix the currents on the spanning tree edges since probabilities can  temporarily accumulate on the vertices (Kirchoff's current law does not apply at any instant of time). However, $\overline{\mathcal{J}^{ps}}$ on the fundamental edges do dictate $\overline{\mathcal{J}^{ps}}$ on the spanning tree, since the average probabilities are conserved.

\section{Mapping SP to NESS}\label{Sec:FromSP2NESS}

In this section we consider the inverse problem defined at the end of Sec.~\ref{Subsec:Defs}, which conceptually is the simpler direction: given a time-dependent periodic rate matrix $\mathcal{W}(t)$, how do we construct a time-independent rate matrix $\mathcal{R}$ whose steady state properties satisfy Eq.(\ref{eq:goal})?

To construct $\mathcal{R}$, we first solve for the periodic averages  $\overline{p_{i}^{ps}}$, $\overline{\mathcal{J}_{ij}^{ps}}$ and $\overline{\sigma^{ps}_{ij}}$ associated with $\mathcal{W}(t)$. These can be calculated by obtaining the periodic solution of the master equation, $p^{ps}_i(t)$, and then plugging this solution into Eq.(\ref{Eq:TimeVaerageProb}). In most cases, however, finding the periodic solution must be done numerically.

Once $\overline{p_{i}^{ps}}$, $\overline{\mathcal{J}_{ij}^{ps}}$ and $\overline{\sigma^{ps}_{ij}}$ are known, we next have to build a rate matrix $\mathcal{R}$ whose steady state properties, $ \vec p^{\,ss}$, $\mathcal{J}^{ss}$ and $\sigma^{ss}$, satisfy Eq.(\ref{eq:goal}). A nice consequence of Eq.(\ref{Eq:R_decomposition}) is that if all the currents along the graph edges are non-zero, then {\it the quantities $ \vec p^{\,ss}$, $\mathcal{J}^{ss}$ and $\sigma^{ss}$ uniquely determine $\mathcal{R}$}. We express this relationship by the shorthand notation
\begin{equation}
\label{eq:uniqueR}
\{  \vec p^{\,ss} , \mathcal{J}^{ss} , \sigma^{ss} \} \Rightarrow \mathcal{R}
\end{equation}
To see this, we use Eq.(\ref{Eq:R_decomposition}) to write the entropy production rates as 
\begin{eqnarray}\label{Eq:sigma_simp}
\sigma_{ij}^{ss} = \mathcal{J}_{ij}^{ss}\log\frac{\mathcal{S}_{ij} + \frac{1}{2}\mathcal{J}^{ss}_{ij}}{\mathcal{S}_{ij} - \frac{1}{2}\mathcal{J}^{ss}_{ij}}.
\end{eqnarray}
By this equation, the elements of $\mathcal{J}^{ss}$ and $\sigma^{ss}$ uniquely determine $\mathcal{S}$, if the currents are non-zero:
\begin{equation}
\{  \mathcal{J}^{ss} , \sigma^{ss} \} \Rightarrow \mathcal{S}
\end{equation}
If some of the currents are zero, then the corresponding elements of $\mathcal{S}$ are not uniquely determined from the currents and entropy production alone, and they can be arbitrarily chosen. Once $\mathcal{S}$ has been obtained consistently with $\mathcal{J}^{ss}$ and $\sigma^{ss}$, it can be combined with $\mathcal{P}$ and $\mathcal{J}^{ss}$, via Eq.(\ref{Eq:R_decomposition}), to give  $\mathcal{R}$.

In the remainder of this paper, we will address the forward problem,
namely how to construct, for a given NESS, a mimicking SP protocol. While this problem is conceptually more complicated than the inverse problem discussed above, it turns out that it is computationally much simpler and does not require any solution of differential equations.  
\begin{figure}
\centering
\includegraphics[width=0.7\linewidth]{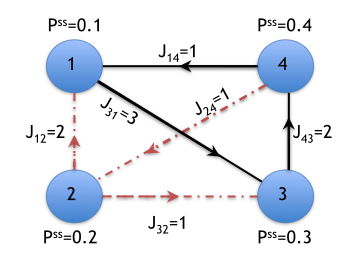}
\caption{A concrete example of a NESS for which we build an equivalent stochastic pump. In this example $\mathcal{R}_{ij}\neq 0$ for any $i,j$. The spanning tree was chosen to be the 2-1, 2-3 and 2-4 edges (dashed red lines), and the fundamental currents are the currents along the 1-4, 1-3 and 3-4 edges (solid black lines). In this system, there are two current-loops: $1\rightarrow 3\rightarrow 4\rightarrow 1$ and $2\rightarrow 3\rightarrow 4\rightarrow2$.}
\label{fig:Figure_examp_system_NESS}
\end{figure}

\section{Key Ideas }\label{Sec:KeyIdea}
Here we establish three technical results that will play a crucial role in the construction of $\mathcal{W}(t)$.

\subsection{No Current-Loops in Detail Balance Systems}\label{Sec:NoLoops}
A system satisfying detailed balance has non-vanishing currents
when the instantaneously probability distribution differs from the equilibrium state of the instantaneous rate matrix. These currents, however, cannot form a {\it current loop}.
That is, no loop $i,j,k,\cdots m,i$ on the graph associated with the system can have all the currents oriented in the same direction around the loop. As an example of a current loop, consider the system described in Fig. (\ref{fig:Figure_examp_system_NESS}). No detailed balance system can have instantaneous currents equal to the currents in the loop $1\rightarrow 3\rightarrow 4\rightarrow 1$ (or in the loop $2\rightarrow 3\rightarrow 4\rightarrow2$), since they all have the same orientation.  

To see why a current loop is inconsistent with detailed balance, let us consider the currents along the edges of a loop $i,j,k,...m,i$. Using Eq.(\ref{Eq:R_decomposition}) to decompose the detailed balance rate matrix  $\mathcal{W} =\mathcal{S}{\Pi}^{-1}$, the currents generated by a distribution $\vec q$ satisfy
\begin{eqnarray}
\frac{\mathcal{J}_{ij}}{\mathcal{S}_{ij}} &=& \pi_j^{-1}q_j - \pi_i^{-1}q_i,\\
\frac{\mathcal{J}_{jk}}{\mathcal{S}_{jk}} &=& \pi_k^{-1}q_k - \pi_j^{-1}q_j,\\
\vdots\nonumber\\
\frac{\mathcal{J}_{mi}}{\mathcal{S}_{mi}} &=& \pi_i^{-1}q_i - \pi_m^{-1}q_m.
\end{eqnarray}     
Summing these equations we get that $\sum (\mathcal{J} / \mathcal{S}) =0$ around the loop. This means that not all the currents can have the same sign, since each $\mathcal{S}_{ij}>0$. Therefore there are no current loops. (A similar argument was used in \cite{mandal2011No_pump_graphical}).

By contrast, the steady state currents of a NESS must form at least one current-loop. To see this, just choose a site (denoted by $i$) through which some of the steady state currents flow. Conservation of probability implies that at least one of these currents is going out of the site $i$, so we choose such a current say from $i$ into site $j$. Now from the site $j$ again, there is at least one current going out, say to to site $k$. Following the same argument, from any site we can ``flow'' with a current into a new site, but since the number of sites is finite, after no more than $n$ such steps we must come back to a site we already visited. Therefore, there must exist at least one current-loop.

\begin{figure}
	\centering
	\includegraphics[width=0.9\linewidth]{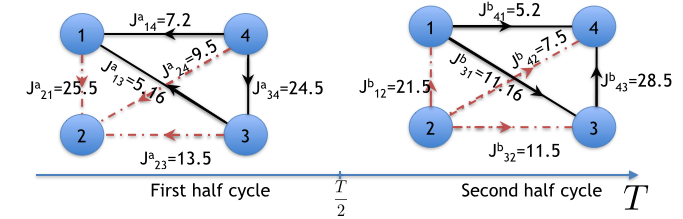}
	\caption{ The currents of the stochastic pump constructed to mimic the NESS in Fig.(\ref{fig:Figure_examp_system_NESS}). The pumped currents during the first (left) and second (right) half-periods are shown. On each of the edges, the current directions in the two half periods are opposite. Note that there are no current-loops in both of the half periods.  }
	\label{fig:Examp_SP}
\end{figure}

\subsection{Transformation for the diagonal part of the decomposition of $\mathcal{W}$}\label{Sec:transformation_diag}

Suppose  we have a detailed balance rate matrix $ \hat{\mathcal{W}}$ and two instantaneous probability distributions $\vec q$ and $\vec p$, neither of which necessarily correspond to a stationary distribution.
We would like to transform $\hat{\mathcal{W}}$ into a different detailed balance rate matrix, ${\mathcal{W}}$, such that $\hat{\mathcal{W}}\vec q = \mathcal{W}\vec p$. This is achieved by the following transformation
\begin{subequations}
\label{eq:trans}
\begin{eqnarray}\label{Eq:Transformation}
	\mathcal{W} =\hat{\mathcal{W}}\mathcal{Q}\mathcal{P}^{-1},
\end{eqnarray} 
where $\mathcal{P}$ and $\mathcal{Q}$ are the diagonal matrices corresponding to $\vec p$ and  $\vec q$, respectively. Using indices, this reads 
\begin{eqnarray}\label{Eq:Trans_components}
\mathcal{W}_{ij} = \hat{\mathcal{W}}_{ij}q_jp_j^{-1}.
\end{eqnarray}
\end{subequations}
The transformation given by Eq.(\ref{Eq:Transformation}) affects only the diagonal part of the decomposition and has the following properties: 
\begin{enumerate}
	\item If $\hat{\mathcal{W}}$ satisfies the detailed balance condition, then so does ${\mathcal{W}}$. This can be seen by using the decomposition of $\hat{\mathcal{W}}$ as in Eq.(\ref{Eq:R_decomposition}) in the above transformation and noting that $\mathcal{W}$ is a symmetric rate matrix times a diagonal matrix, and therefore it has no currents in its steady state. 
	\item The instantaneous currents of a system described by $\hat{\mathcal{W}}$ with probabilities $\vec q$ are the same as those of a system described by ${\mathcal{W}}$ with the probabilities $\vec p$. This follows by substituting Eq.(\ref{Eq:Trans_components}) into Eq.(\ref{Eq:PerCurrentsDef}).
	\item From Eqs.(\ref{Eq:Trans_components},\ref{eq:PerSigmaDef}) it follows that the instantaneous entropy production rates along each edge ($\sigma_{ij}$) for a system described by $\hat{\mathcal{W}}$ with probabilities $\vec q$ and for a system described by ${\mathcal{W}}$ with probabilities $\vec p$ are the same.  
\end{enumerate}

The significance of this transformation can be stated as follows.
If we have a rate matrix $\hat{\mathcal{W}}$ and probabilities $\vec q$, which produce instantaneous currents $\mathcal{J}$ and entropy production rates $\sigma$, then for any other probability distribution $\vec p$ we can construct the rate matrix $\mathcal{W}$ that generates the same $\mathcal{J}$ and $\sigma$.

\subsection{Transformation for the symmetric part of the decomposition of $\mathcal{W}$}\label{Sec:TransfCurrents}
Currents arise in a system described by a detailed balance $\mathcal{W}$ when the instantaneous probability distribution $\vec p$ differs from the equilibrium distribution, $\vec\pi$. 
We next show how to vary the magnitudes of these currents  (but not their directions) while keeping $\vec p$ and $\vec{\pi}$ fixed. As the directions of the currents do not vary under this transformation, no loops can be formed in accordance with the ``No loop condition'' in Sec.(\ref{Sec:NoLoops}). 

Let us use the decomposition $\mathcal{W} = \mathcal{S}\Pi^{-1}$ where $\mathcal{S}$ is symmetric and $\Pi = {\rm diag}(\vec\pi)$. The currents are then given by 
\begin{equation}\label{Eq:Currents_decomp}
\mathcal{J}_{ij} = \mathcal{S}_{ij}\left(\pi_i^{-1}p_j - \pi_j^{-1}p_i\right)
\end{equation}
We see that by varying $\mathcal{S}_{ij}$ we vary the magnitude of the current $\mathcal{J}_{ij}$, but not its sign, since $\mathcal{S}_{ij}\ge 0$.

This transformation is complementary to the one given by Eq.(\ref{eq:trans}): it enables us to change the currents (and entropy production rates) while keeping the probabilities $\vec p$ and $\vec\pi$ fixed, by tuning the symmetric part of $\mathcal{W}$ in the decomposition given by Eq.(\ref{Eq:R_decomposition}). 
By contrast, with the previous transformation we can vary the probabilities  $\vec p$ and $\vec\pi$ at fixed currents and entropy production rates, by tuning the diagonal part of $\mathcal{W}$.

\section{Construction of the pumping protocol}\label{Sec:Construction}
Given a rate matrix $\mathcal{R}$ one can calculate its steady state $ \vec p^{\,ss}$ (the null eigenvector of $\mathcal{R})$ and thus $\mathcal{J}^{ss}$ and $\sigma^{ss}$ using Eqs.(\ref{Eq:CurrentsDef},\ref{eq:SigmaDef}). Our goal is to construct a periodic pumping protocol -- a time-dependent detailed balance rate matrix $\mathcal{W}(t)$ with some period $T$ -- such that Eqs.(\ref{eq:goal}) hold, or in other words a SP that mimics the NESS. For simplicity, we assume that there are no edges along which the steady state currents are zero. The case involving zero currents along some edges is analyzed in the appendix. 

We note that in general there are many SP's that mimic any specific NESS -- the mapping is not one-to-one. Out of the many SP that mimic the NESS, we wish to choose one using a relatively simple construction, yet generic enough to mimic any NESS.  Naively we would like to have a  SP whose periodic state gives rise to time-independent quantities, $p_i^{ps}(t) = p_i^{ss}, \, \mathcal{J}_{ij}^{ps}(t) = \mathcal{J}_{ij}^{ss}$ and $\sigma_{ij}^{ps}(t) = \sigma_{ij}^{ss}$. Such a construction is, unfortunately, impossible. This can be seen from the result of Sec. (\ref{Sec:NoLoops}), which states that for all $t$, $\mathcal{J}_{ij}^{ps}(t)$ cannot have any current loops, whereas $\mathcal{J}_{ij}^{ss}$ must have at least one current loop.  Thus, we can only hope to achieve a mapping between the {\it time-averaged} quantities associated with the SP and those of the NESS, as in Eq.(\ref{eq:goal}). This also implies that at least some of the currents of the SP must be time-dependent. 

The construction described below, though the simplest we could find, is nevertheless somewhat convoluted. We therefore first give an overview before proceeding to the detailed description. The main reason for the complication is the fact that the periodic solution, $p_i^{ps}(t)$, is a highly non-trivial function of the pumping protocol $\mathcal{W}(t)$. To avoid this complication, we construct simultaneously both the pumping protocol $\mathcal{W}(t)$ and its periodic solution $p_i^{ps}(t)$. This is achieved in 5 steps. 

In the first step, described in Sec.(\ref{Sec:Step1}), we  divide the pumping protocol temporal interval of duration $T$ into two equal half-intervals, designated as $a$ and $b$.
We then assign an arbitrary, fixed detailed balanced matrix $\tilde{\mathcal{W}}^a$ and an arbitrary, fixed probability distribution $\vec q^a$, to be associated with the first half-period of driving.
We will refer to $\tilde{\mathcal{W}}^a$ and $\vec q^a$ together as the {\it seed} for that half-cycle, and these quantities will be used to set the current directions during that time interval.
Next, the seed and current directions for the first half-cycle are used to assign a seed ($\tilde{\mathcal{W}}^b$, $\vec q^b$) and current directions for the second half-cycle.
The current directions during the first half-cycle are opposite of those of the second half-cycle, and both sets are, by construction, consistent with the ``no current loop'' condition in Sec.~(\ref{Sec:NoLoops}).

In the next two steps, we use the seeds to construct fixed sets of currents ($\mathcal{J}_{ij}^a$ and $\mathcal{J}_{ij}^b$) and entropy production rates ($\sigma_{ij}^a$ and $\sigma_{ij}^b$) for the first and second halves of the cycle, whose averages over the two halves are equal to the steady-state values that we wish to mimic:
\begin{equation}
\mathcal{J}_{ij}^{ss} = \frac{1}{2} \left( \mathcal{J}_{ij}^a + \mathcal{J}_{ij}^b \right)
\quad,\quad
\sigma_{ij}^{ss} = \frac{1}{2} \left( \sigma_{ij}^a + \sigma_{ij}^b \right)
\end{equation}
for all $i\ne j$.
This is done first for the fundamental currents in Sec. (\ref{Sec:Step2}) and then for the spanning tree currents in Sec.(\ref{Sec:Step3}).

Up to this point, the currents and entropy production rates for the two half-cycles have been constructed from the initial, arbitrary seeds, but the corresponding time-periodic rate matrices $\mathcal{W}^a(t)$ and $\mathcal{W}^b(t)$ that actually generate these currents and entropies are not yet known.
In the fourth step, described in Sec. (\ref{Sec:Step4}), we use the transformation of Sec. (\ref{Sec:TransfCurrents}) to adjust the symmetric part of the seed matrices, $\tilde{\mathcal W}^{a,b}$, arriving at new rate matrices $\hat{\mathcal W}^{a,b}$ that produce the desired currents ${\mathcal J}^{a,b}$, for the seed probabilities $\vec q^{a,b}$.
These currents, together with the desired averaged probabilities $\overline{p_i^{ps}}$, fix $p_i^{ps}(t)$. 
In the last step (Sec.~\ref{Sec:Step5}) we use the transformation of Sec.~(\ref{Sec:transformation_diag}), together with the symmetric parts of $\hat{\mathcal W}^{a,b}$, to construct rate matrices $\mathcal{W}^a(t)$ and $\mathcal{W}^b(t)$ for which $p_i^{ps}(t)$ is the periodic solution of the master equation.

In the specific protocol described below, the entries of the matrix $\mathcal{W}(t)$ are not continuous functions of time, as they have discontinuities between the two $T/2$ intervals. These discontinuities are not essential, and can be removed at the expense of making the construction less transparent.

To improve the clarity of presentation, some of the formal definitions of the construction are followed by a concrete application to the example of a 4-state system described in Fig.(\ref{fig:Figure_examp_system_NESS}). In this example the NESS system has 4 states with $ \vec p^{\,ss} = (0.1,0.2,0.3,0.4)$. The fundamental currents were chosen such that they form a loop, and their values are $\mathcal{J}_{31}^{ss}=3$, $\mathcal{J}_{43}^{ss} = 2$ and $\mathcal{J}_{14}^{ss}=1$. The currents for the spanning tree edges are then dictated by Kirchoff's law - the sum of currents in each vertex must be zero. The corresponding current-matrix is
\begin{eqnarray}
\mathcal{J}^{ss} = \left(\begin{array}{cccc}
0 & 2&-3 & 1\\
-2& 0& 1& 1\\
3 & -1& 0& -2\\
-1& -1& 2& 0
\end{array}\right)
\end{eqnarray}
Finally, we choose the entropy production rate to be 1 along all the edges. Using Eq.(\ref{Eq:sigma_simp}) the matrix $\mathcal{S}$ can be  calculated, and is given in Eq.(\ref{Eq:S_mat_example}) in the Appendix.
The matrix $\mathcal{R}$ giving rise to this particular NESS can be constructed using Eq.(\ref{Eq:R_decomposition}).

\subsection{Step 1- choosing the seed}\label{Sec:Step1}

In what follows, superscripts $a$ and $b$ stand for quantities associated with the first and second halves of the period, respectively.

Our first step is to choose an arbitrary detailed balance matrix on the graph. This is done by choosing an equilibrium state for the first half period $\vec \pi^{a}$  and a symmetric rate matrix $\tilde{\mathcal{S}}$  from which we compose $\tilde{\mathcal{W}}^a = \tilde {\mathcal{S}}({\Pi}^{a})^{-1}$ where $\Pi^a = {\rm diag}(\vec\pi^a)$.  Next, we choose a fixed probability distribution $\vec q^a \ne \vec\pi^a$ that satisfies
\begin{eqnarray}\label{Eq:ConditionForEntropy}
\left|\log\frac{\pi_i^aq^a_j}{q^a_i\pi_j^{a}}\right|<\left|\log\frac{\mathcal{R}_{ij}p^{ss}_j}{\mathcal{R}_{ij}p^{ss}_i}\right|
\end{eqnarray}
for any $i \ne j$.
It is always possible to satisfy this condition, by choosing $\vec q^a$ close enough to $\vec{\pi}^a$.
As we will see in the next section, this condition is necessary for the consistency of our construction. 

For the second half of the period we replace $\vec \pi^{a}$ and $\vec q^a$ by vectors with components $\pi^{b}_i ={1}/\pi^{a}_i$ -- from which we construct $\tilde{\mathcal{W}}^b= \tilde {\mathcal{S}}({\Pi}^{b})^{-1}$ -- and $q^b_i = {1}/q^a_i$. The two probability vectors $\vec \pi^{b}$ and $\vec q^b$ are not normalized, but as will become clear, this normalization does not play any role, and it will prove to be simpler to work with these unnormalized vectors. We note that the currents generated by  $\tilde{\mathcal{W}}^a$ and $\vec q^a$,
\begin{equation}
\tilde{\mathcal{J}}_{ij}^a = \tilde{\mathcal{S}}_{ij}\left(\pi^{a}_jq^a_j - \pi^{a}_iq^a_i\right),
\end{equation}
have, on each edge, opposite signs to those generated by $\tilde{\mathcal{W}}^b$ and $\vec q^b$, given by 
\begin{equation}
\tilde{\mathcal{J}}_{ij}^b = \tilde{\mathcal{S}}_{ij}\left( \frac{1}{\pi^{a}_jq^a_j} - \frac{1}{\pi^{a}_iq^a_i}\right) .
\end{equation}
The values of  $\tilde{\mathcal{J}}_{ij}^{a,b}$ will not explicitly be used in what follows -- only their directions, which by construction are consistent with detail balance.
We additionally note that
\begin{equation}
\log\frac{\pi_i^aq^a_j}{q^a_i\pi_j^{a}} = -\log\frac{\pi_i^bq^b_j}{q^b_i\pi_j^{b}} \quad.
\end{equation}

To illustrate this part of the construction with our four-state example, we choose
\begin{eqnarray}
\tilde{\mathcal{S}} = \left(\begin{array}{cccc}
-3& 1 &  1&  1\\ 
1& -3 &  1& 1 \\ 
1& 1 &  -3& 1 \\ 
1&  1&1  & -3
\end{array} \right) \quad,\quad \vec{\pi}^a = \left(\begin{array}{c}
1/4\\ 
1/4\\ 
1/4\\ 
1/4
\end{array} \right)
\end{eqnarray}
Both $\tilde{\mathcal{S}}$ and $\vec \pi^a$ are, in fact, arbitrary.

We next note that $\min_{ij}[\log{\mathcal{R}_{ij}p^{ss}_j}-\log{\mathcal{R}_{ji}p^{ss}_i}] = 1/3$.  If we therefore choose $\vec q^a$ such that $\left|\log{\pi_i^a}-\log{q^a_i}\right|<1/6$,then Eq.(\ref{Eq:ConditionForEntropy}) is satisfied. Any vector close enough to $\vec\pi^a$ will do. As an example we use $\vec q^a = (0.23,0.24,0.26,0.27)$. In the second half of the period these correspond to $\vec{\pi^b} = (4,4,4,4)$ and $\vec q^b =(4.3478,4.1667,3.8462,3.7037)$.

\subsection{Step 2 - fundamental edge currents}\label{Sec:Step2}
In this step we set the currents along the fundamental edges to be constant during each of the two half cycles, such that their time averages  (the average between the first and second halves) is equal to $\mathcal{J}^{ss}$, and the time average of the entropy production rates is equal to $\sigma^{ss}$.  Moreover, we will choose these fundamental currents to have the same directions as the fundamental currents in $\tilde{\mathcal{J}}^a$ during the first half-cycle, and to be reversed in direction during the second half-cycle.

We first make sure that for each fundamental edge, the average of $\mathcal{J}_{ij}^a $ and $\mathcal{J}_{ij}^b$, and therefore the time averaged current, is exactly $\mathcal{J}_{ij}^{ss}$.
To that effect we introduce the following rule:
 \begin{itemize}
 	\item If the direction of the NESS current, $\mathcal{J}_{ij}^{ss}$, is the same as the direction of $\tilde{\mathcal{J}}_{ij}^a$ defined above, then we set in the first half of the period $\mathcal{J}_{ij}^a = (2+\alpha_{ij})\mathcal{J}_{ij}^{ss} $ and in the second half of the period  $\mathcal{J}_{ij}^b = -\alpha_{ij}\mathcal{J}_{ij}^{ss} $, where $\alpha_{ij}$ are positive and  will be determined below.
 	\item If the direction of the NESS current, $\mathcal{J}_{ij}^{ss}$, is not the same as the direction of $\tilde{\mathcal{J}}_{ij}^a$, then we set in the first half of the period $\mathcal{J}_{ij}^a = -\alpha_{ij}\mathcal{J}_{ij}^{ss} $ and in the second half   $\mathcal{J}_{ij}^b = (2+\alpha_{ij})\mathcal{J}_{ij}^{ss} $. 
 \end{itemize}
According to this rule, the directions of the currents during the first half of the period are the same as that of $\tilde{\mathcal{J}}_{ij}^a $ and are opposite to those in the second half.
Moreover, by the above construction the time averaged currents on the fundamental edges have the required values:
\begin{equation}
\frac{1}{2} \left( \mathcal{J}_{ij}^a + \mathcal{J}_{ij}^b \right) = \mathcal{J}_{ij}^{ss}.
\end{equation}

Next we determine the values of the $\alpha_{ij}$'s so as to satisfy the requirement on the entropy production rates.  Assuming for the moment that the probability distributions in the first and second halves of the period are given by $\vec q^a$ and $\vec q^b$ and that the equilibrium distributions of the detailed balance matrices during the first and second halves of the period are given by $\vec \pi^a$ and $\vec \pi^b$ respectively, then the entropy production rates with the currents $\mathcal{J}_{ij}^a $ and $\mathcal{J}_{ij}^b$ during the two halves of the cycle are given by 
	\begin{eqnarray}
	\sigma^{a}_{ij} = \mathcal{J}_{ij}^{a}\log\frac{\pi_j^{a}q_j^{a}}{\pi_i^{a}q_i^{a}}\nonumber\\
	\sigma^{b}_{ij} = \mathcal{J}_{ij}^{b}\log\frac{\pi_j^{b}q_j^{b}}{\pi_i^{b}q_i^{a}}.
	\end{eqnarray}
	Substituting in these equations  $\mathcal{J}_{ij}^{a}$ and $\mathcal{J}_{ij}^{b}$ in terms of $\alpha_{ij}$ and demanding that  $\frac{1}{2}(\sigma_{ij}^a + \sigma_{ij}^a) =\sigma^{ss}_{ij}$, gives an equation for $\alpha_{ij}$ whose solution is
 \begin{eqnarray}\label{Eq:alpha_ij}
 \alpha_{ij} = \left|\left(\log\frac{\pi_i^aq^a_j}{q^a_i\pi_j^{a}}\right)^{-1}\log\frac{\mathcal{R}_{ij}p^{ss}_j}{\mathcal{R}_{ij}p^{ss}_i} \right|-1.
 \end{eqnarray}  
Eq.(\ref{Eq:ConditionForEntropy}) ensures that indeed $\alpha_{ij}>0$.

To illustrate this step, we examine the signs of the currents generated by the  matrix $\tilde{\mathcal{W}}^a = \tilde{\mathcal{S}}(\Pi^a)^{-1}$ and the probability $\vec q^a$, denoted by $\tilde{\mathcal{J}}^a$, on the fundamental edges:
\begin{eqnarray}
sign\left(\tilde{\mathcal{J}}^a_{13}\right) = +1, &
sign\left(\tilde{\mathcal{J}}^a_{14}\right) = +1, &
sign\left(\tilde{\mathcal{J}}^a_{34}\right) = +1.
\end{eqnarray}
While the direction of the current along the 1-4 edge is the same as the current orientation of $\mathcal{J}^{ss}$, for the other two fundamental edges the directions of $\mathcal{J}^a$ and $\mathcal{J}^{ss}$ are not the same.

We next solve Eq.(\ref{Eq:alpha_ij}) for $\alpha_{ij}$. The explicit expression, as well as the currents, are given in the appendix. Fig(\ref{fig:Examp_SP}) shows the currents on the first half period (left) and second half period (right). Note that (i) the direction of the currents along each edge are opposite in the two halves of the period (ii) as discussed above, in the first half period the direction of the current along the 1-4 edge is the same as that of Fig(\ref{fig:Figure_examp_system_NESS}), but the direction of currents along the other fundamental edges are different from that of Fig(\ref{fig:Figure_examp_system_NESS}),  and (iii) there are no current loops in Fig(\ref{fig:Examp_SP}).

 \subsection{Step 3 - Spanning tree  currents}\label{Sec:Step3}
So far we have shown how to construct the fundamental edge currents. Next we discuss the currents along the edges of the spanning tree. For both the first and the second half-periods we impose the following two constraints:
 \begin{enumerate}
 	\item The sum of currents feeding into any site $i$ during the first half-period must be equal to minus the same quantity during the next half-period:  
 	\begin{eqnarray}\label{Eq:SpanTreeCond1}
 	\sum_j\mathcal{J}^a_{ij} = - \sum_j \mathcal{J}^b_{ij}
 	\end{eqnarray}
 	These constraints ensure that  we indeed have a periodic time evolution: $\Delta p_i = \int_0^T \partial_tp_idt = (T/2)\sum_j (\mathcal{J}^a_{ij} + \mathcal{J}^b_{ij}) = 0$. Note that these are only $n-1$ independent equations, since conservation of probability adds the constraint $\sum_{ij} \mathcal{J}_{ij}^{a,b} =  0$ .
 	\item For each spanning tree edge we demand that 
 	\begin{eqnarray}\label{Eq:SpanTreeCond2}
 	\mathcal{J}_{ij}^a\log\frac{\pi_i^aq^a_j}{q^a_i\pi_j^{a}} + \mathcal{J}_{ij}^b\log\frac{\pi_i^bq^b_j}{q^b_i\pi_j^{b}}  = 2\sigma_{ij}^{ss}.
 	\end{eqnarray}  
 	The number of edges in the spanning tree is $n-1$, and therefore these are $n-1$ additional equations.  They ensure that the time-averaged entropy production rate along the $i,j$ edge are the same as $\sigma_{ij}^{ss}$.
 \end{enumerate}   
 All together Eqs.~(\ref{Eq:SpanTreeCond1}, \ref{Eq:SpanTreeCond2}) are $2(n-1)$ linear equations for $2(n-1)$ unknowns: $\mathcal{J}^{a,b}$ on the spanning tree, which has $n-1$ edges. Moreover, the directions  $\mathcal{J}_{ij}^{a,b}$ that solve these equations  are the same as that of $\tilde{\mathcal{J}}_{ij}^{a,b}$ respectively. To see this, let us use the definition of $\vec q^b$ and $\vec\pi^b$ in the second condition above, together with the definition of $\sigma^{ss}_{ij}$  (Eq.\ref{eq:SigmaDef}):
 \begin{eqnarray}
 \left(\mathcal{J}_{ij}^a - \mathcal{J}_{ij}^b\right) =2\mathcal{J}_{ij}^{ss} \left(\log\frac{\pi_i^aq^a_j}{q^a_i\pi_j^{a}}\right)^{-1}\log\frac{\mathcal{R}_{ij}p^{ss}_j}{\mathcal{R}_{ji}p^{ss}_i}.
 \end{eqnarray}
 Taking the absolute value of both sides in the above equation and using Eq.(\ref{Eq:ConditionForEntropy}) implies that 
 \begin{eqnarray}\label{Eq:Currents_Diff}
 \left|\mathcal{J}_{ij}^a - \mathcal{J}_{ij}^b\right|>2\left|\mathcal{J}_{ij}^{ss}\right|.
 \end{eqnarray} 
 However, by the construction of $\mathcal{J}_{ij}^a$ and $\mathcal{J}_{ij}^b$ their average is equal to the steady state current, $\mathcal{J}_{ij}^{ss}$, therefore
 \begin{eqnarray}\label{Eq:CurrentAvg}
 \left(\mathcal{J}_{ij}^a + \mathcal{J}_{ij}^b\right)/2=\mathcal{J}_{ij}^{ss}.
 \end{eqnarray}  
 Eqs.(\ref{Eq:CurrentAvg}, \ref{Eq:Currents_Diff}) are consistent with each other only if the sign of $\mathcal{J}_{ij}^a$ is opposite to that of $\mathcal{J}_{ij}^b$. Moreover, as $\sigma_{ij}^{ss}>0$ (this follows from Eq.(\ref{Eq:sigma_simp})), the signs of $\mathcal{J}_{ij}^a$ and $\mathcal{J}_{ij}^b$ must be the same as that of $\log{\pi_i^aq^a_j/q^a_i\pi_j^{a}}$ and $\log{\pi_i^bq^b_j/q^b_i\pi_j^{b}}$, respectively, otherwise the left hand side of Eq.(\ref{Eq:SpanTreeCond2}) will be negative. Therefore it is also the same as the sign of $\tilde{\mathcal{J}}_{ij}^a$.

 \subsection{Step 4 - Transforming the symmetric part  for the currents} \label{Sec:Step4}
 At this stage we have constructed the currents for both half cycles with the same directions as the currents of $\tilde{\mathcal{W}}^{a,b}$ with $\vec q^{a,b}$. We can now use the transformation of the symmetric part of the decomposition of the rate matrix (defined in section \ref{Sec:TransfCurrents}) to change the symmetric part of $\tilde{\mathcal{W}}$ such that the currents generated by the transformed matrix and the seed probabilities $\vec{q}^{a,b}$ have the constructed values:
 \begin{eqnarray}\label{Eq:equations_for_S}
 \mathcal{S}^a_{ij} &=& \frac{\mathcal{J}^a_{ij}}{(\pi^a_j)^{-1} q^a_j - (\pi^a_i)^{-1} q^a_i}\label{Eq:Def_S1}\nonumber,\\
 \mathcal{S}^b_{ij} &=& \frac{\mathcal{J}^b_{ij}}{(\pi^b_j)^{-1} q^b_j - (\pi^b_i)^{-1} q^b_i}\label{Eq:Def_S2}.
  \end{eqnarray} 
 Importantly, all these off-diagonal elements are positive. This follows from the fact that the denominators are just $\tilde{\mathcal{J}}_{ij}^{a,b}/\tilde{\mathcal{S}}_{ij}$, but by our construction the signs of $\mathcal{J}^{a,b}_{ij}$ is the same as that of $\tilde{\mathcal{J}}_{ij}^{a,b}$. The diagonal elements of $\mathcal{S}^{a,b}$ are now determined by the requirement that the sum of each of the columns is zero.

 \subsection{Step 5 - Forming a solution to the master equation} \label{Sec:Step5}
 We have now arrived at the matrices $\hat{\mathcal{W}}^{a,b} = \mathcal{S}^{a,b}(\Pi^{a,b})^{-1}$.
 These are the transformed seed matrices, which have been constructed (by tuning the elements of $\mathcal{S}^{a,b}$) so as to produce the desired currents and entropy production rates when the probability vectors are $\vec q^{a,b}$. However, the time average of $\vec q^{a,b}$ is not $\vec{p}^{ss}$, and in fact $\vec q^{a,b}$ are not solutions of the master equation: $\partial_t{\vec{q}}^{a,b}\neq \hat{\mathcal{W}}^{a,b}\vec{q}^{a,b}$ . To remedy this situation we use the transformation of the diagonal part of the decomposition of $\hat{\mathcal{W}}$, defined in section \ref{Sec:transformation_diag}.
 
 First, we want the time-averaged probabilities to be equal to $ \vec p^{\,ss}$. Second, we already know what $\partial_t p_i$ should be, in terms of the desired currents $\mathcal{J}^{a,b}$.
 Namely, $\partial_t p_i = \sum_j\mathcal{J}_{ij}$. Therefore we do the following:
  \begin{enumerate}
  	\item Calculate $\vec m^{a,b} =\partial_t\vec p = \hat{\mathcal{W}}^{a,b} \vec q^{a,b}$. These are the temporal slopes of the probabilities that solve the master equation, during the first and second halves of the period. 
  	\item Choose $T$ such that $0<p_i^{ss}\pm(T/4)m_i^{a,b}<1$ for any $i$. This choice ensures that the probabilities stay bounded between 0 and 1, and is always possible by taking $T$ to be small enough.
  	\item Construct $p^{a}_i(t) = p_i^{ss} - (T/4)m_i^{a} + m_i^{a}t$ and $p^{b}_i(t) = p_i^{ss} - (3T/4)m_i^{b} + m_i^{b}t$. These are  the solutions of the master equation in the first and second halves of the period, respectively. 
It follows from Eq.(\ref{Eq:SpanTreeCond1}) that $m_i^b = - m_i^a$. This further implies that the probabilities defined above are continuous functions of time, i.e. $p_i^a(T/2) = p_i^b(T/2)$ and $p_i^a(0)=p_i^b(T)$, as they should be.
  	\item From the above we define: 
  	\begin{eqnarray}\label{Eq:W_final}
  	\mathcal{W}(t) = \begin{cases}
  	\mathcal{S}^a(\Pi^a)^{-1}\mathcal{Q}^a(\mathcal{P}^a)^{-1}(t) \quad &,\quad t< T/2 \\
    \mathcal{S}^b(\Pi^b)^{-1}\mathcal{Q}^b(\mathcal{P}^b)^{-1}(t) \quad &,\quad t> T/2
      	\end{cases}. 
  	\end{eqnarray}
  \end{enumerate} 
The matrix $\mathcal{W}(t)$ has all the periodic state averages we demand, and its periodic state solutions in the two halves of the period are, by construction, $p^{a,b}(t)$.

For our example, we calculate the slopes $\partial_t\vec p$ on the two half cycles by proper summation of the currents: $\partial_t\vec p^a = (37.89,-2.51,5.85,-41.22 )$ and $\partial_t\vec p^b = (-37.89,2.51,-5.85,41.22 )$, which as expected are equal in magnitude and opposite in sign. 
To satisfy $0<p_i(t)<1$, we must choose $T$ small enough, say  $T = 0.01$.   Using $T$ we can obtain the time-dependent probability distributions $\vec p^{a,b}(t)$. These linear functions are plotted in Fig.(\ref{fig:Solution_vs_Construction_Pt}). Plugging these into Eq.(\ref{Eq:W_final}) gives $\mathcal{W}(t)$. To verify that the solution of the master equation with the constructed $\mathcal{W}(t)$ has the required properties we solve this system numerically.  The numerical results for $\vec p(t)$, shown as red circles in Fig(\ref{fig:Solution_vs_Construction_Pt}), agree (up to numerical error) with the analytical solution.  
\begin{figure}
\centering
\includegraphics[width=0.7\linewidth]{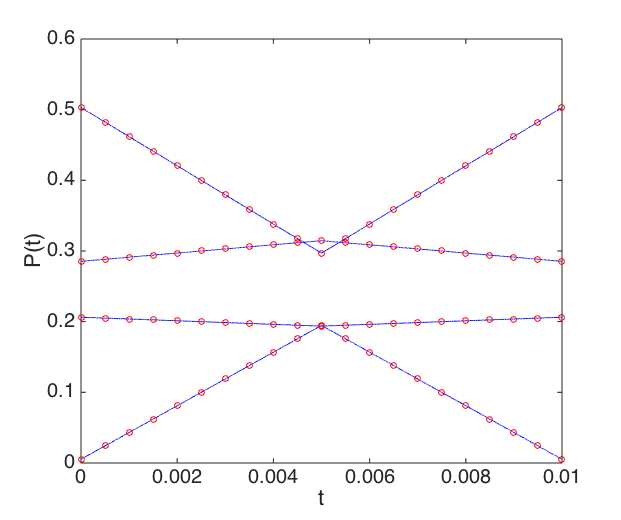}
\caption{$\vec p(t)$ of the proposed construction for our four-state example. The blue line is the exact (constructed) solution, and the red dots represent the numerical solution of the master equation.}
\label{fig:Solution_vs_Construction_Pt}
\end{figure}

\section{Conclusions}\label{Sec:Conclusion}
We conclude with a few comments on our construction.
First, the protocol presented above is clearly not unique. For example, different seeds or choices of a spanning tree result in different protocols. The non-uniqueness might be used, in principle, to match additional quantities, e.g. fluctuations around the average or the rate of decay towards the steady and periodic states. In addition, as mentioned previously, in our construction $\mathcal{W}(t)$ has discontinuities between the two halves of the period. This results from discontinuities in $\mathcal{S}$, $\Pi$ and $\mathcal{Q}$. The discontinuities in $\mathcal{S}$ can be avoided if the currents are not taken to be fixed during the two halves of the period, but changing with time, such that the currents at $t={T}/{2}$ vanish. The discontinuities in $\Pi$ and $\mathcal{Q}$ can be avoided if we change $\vec q$ as a function of time, crossing $\vec{\pi}$ at $t={T}/{2}$. This, however, makes the construction more cumbersome. Next, we note that in the above construction both the symmetric  and the diagonal parts of the decomposition of $\mathcal{W}$ change with time. This is known to be an essential feature of all pumping protocols \cite{rahavNo_pump_PRL}.

Lastly,  we note that for NESS there is no minimal entropy production rate associated with a given set of currents, as is evident from Eq.(\ref{Eq:sigma_simp}): for any $\mathcal{J}$  the entropy production rates of the NESS can be made arbitrarily small by taking $\mathcal{S}$ to be large enough \cite{Schmittmann_zia_2007}. 
The mapping presented here shows this is also the case for stochastic pumps: finite currents can be pumped with arbitrarily small values of dissipation, which is somewhat surprising. Stated more generally, for any connected graph, both a NESS and a SP can be constructed to have any desired set of (time-averaged) probabilities, non-zero currents and positive entropy production rates, provided the currents obey Kirchhoff's law.

Let us now consider the construction of a stochastic pump with small entropy production rates but large currents. For the entropy production rates of the NESS to be very small with non-vanishing currents, $\log\left(\mathcal{R}_{ij}p^{ss}_j/ \mathcal{R}_{ji}p^{ss}_i\right)$ must be very small, say of order $\varepsilon$. Eq.(\ref{Eq:ConditionForEntropy}) then implies that $\log\left(\pi^a_iq^a_j/\pi^a_j q^a_i\right) \sim \mathcal{O}(\varepsilon)$ as well.
Exponentiating this  gives  $\left(\pi^a_iq^a_j/\pi^a_j q^a_i\right) =1 + \mathcal{O}(\varepsilon)$  thus $(\pi^a_j)^{-1} q^a_j - (\pi^a_i)^{-1} q^a_i\sim\mathcal{O}(\varepsilon)$. By Eq.(\ref{Eq:equations_for_S}) this implies $\mathcal{S}\sim\mathcal{O}(\varepsilon^{-1})$. Therefore, in stochastic pumps -- as in NESS -- small entropy production rates with non-vanishing currents come at the cost of large values in the symmetric part of the rate matrix.
But when the elements of $\mathcal{S}$ are large, the corresponding transitions occur very rapidly.  Thus, we obtain finite currents at arbitrarily low dissipation when there is a large separation of timescales, with transitions among the $n$ states of the system -- and therefore relaxation to equilibrium -- occurring much more rapidly than the external driving of parameters.
The greater the separation of timescales, the more the stochastic pump approaches the adiabatic (quasi-static) limit \cite{Adiabatic_Molecular_motors_2007}, in which the system remains in thermal equilibrium at all times and there is no entropy production.

\paragraph{Acknowledgments:} 
C.J. acknowledges financial support from  the U.S. Army Research Office under contract number W911NF-13-1-0390. O.R. acknowledges financial support from the James S. McDonnell Foundation.

\appendix
\section{Details of the specific example:}
The example we consider is shown in Fig(\ref{fig:Figure_examp_system_NESS}). It has 4 states with $ \vec p^{ss} = (0.1,0.2,0.3,0.4)$. The spanning tree edges are the dashed-doted lines (red), and the fundamental current edges are the solid line (black). The fundamental currents were chosen such that they form a loop, and their values are $\mathcal{J}_{31}^{ss}=3$, $\mathcal{J}_{43}^{ss} = 2$ and $\mathcal{J}_{14}^{ss}=1$. Once these currents are chosen, the currents for the spanning tree edges are dictated by Kirchoff's law - the sum of currents in each vertex must be zero. The corresponding currents matrix is
\begin{eqnarray}
\mathcal{J}^{ss} = \left(\begin{array}{cccc}
0 & 2&-3 & 1\\
-2& 0& 1& 1\\
3 & -1& 0& -2\\
-1& -1& 2& 0
\end{array}\right)
\end{eqnarray}
To set  $\mathcal{S}$, we choose the entropy production rate to be 1 along all the edges. Using Eq.(\ref{Eq:sigma_simp}) the matrix $\mathcal{S}$ is given in this case by 
\begin{eqnarray}\label{Eq:S_mat_example}
\mathcal{S} = \left(\begin{array}{cccc}
-14.248& 4.083& 9.0832& 1.082\\
4.083  & -6.2469& 1.082& 1.082\\
9.0832 & 1.0820& -14.2481& 4.083\\
1.082 & 1.082 & 4.083& -6.2469
\end{array}\right)
\end{eqnarray} 
and using Eq.(\ref{Eq:R_decomposition})  $\mathcal{R}$ can be calculated, though we will not need it in what follows.

Let us demonstrate our construction on this NESS. For the first step we choose
 \begin{eqnarray}
\tilde{\mathcal{S}} = \left(\begin{array}{cccc}
-3& 1 &  1&  1\\ 
1& -3 &  1& 1 \\ 
1& 1 &  -3& 1 \\ 
1&  1&1  & -3
\end{array} \right) & \vec{\pi}^a = \left(\begin{array}{c}
\frac{1}{4}\\ 
\frac{1}{4}\\ 
\frac{1}{4}\\ 
\frac{1}{4}
\end{array} \right)
\end{eqnarray}
We next note that $\min_{ij}[\log{\mathcal{R}_{ij}p^{ss}_j}-\log{\mathcal{R}_{ji}p^{ss}_i}] = 1/3$.  If we therefore choose $\vec q^a$ such that $\left|\log{\pi_i^a}-\log{q^a_i}\right|<1/6$, condition Eq.(\ref{Eq:ConditionForEntropy}) is automatically satisfied. Any vector close enough to $\vec\pi^a$ will do. As an example we use $\vec q^a = (0.23,0.24,0.26,0.27)$. In the second half of the period these correspond to $\vec{\pi^b} = (4,4,4,4)$ and $\vec q^b =(4.3478,4.1667,3.8462,3.7037)$. 

The next step requires the signs of the currents generated by $\tilde{\mathcal{R}}^a = \tilde{\mathcal{S}}(\Pi^a)^{-1}$ with the probability $\vec q^a$:
\begin{eqnarray}
sign\left(\tilde{\mathcal{J}}^1_{13}\right) = +1, &
sign\left(\tilde{\mathcal{J}}^1_{14}\right) = +1,&
sign\left(\tilde{\mathcal{J}}^1_{34}\right) = +1.
\end{eqnarray}
Note that the direction of the current in the $1-4$ edge is the same as the desired current orientation, but for the other two edges the directions are different from the desired currents.

In the second step, we solve Eq.(\ref{Eq:alpha_ij}) for $\alpha_{ij}$. This gives:
\begin{eqnarray}
\alpha_{13} &=& 1.7188,\\
\alpha_{14} &=& 5.2366,\\
\alpha_{34} &=& 12.2484. 
\end{eqnarray}
Together with the directions the fundamental currents in the two half cycles are given by
\begin{eqnarray}
\mathcal{J}^1_{13} = 5.156, & \mathcal{J}^1_{14} =  7.2366,& \mathcal{J}^1_{34} =  24.4969,\\
\mathcal{J}^2_{13} = -11.156, & \mathcal{J}^2_{14} =  -5.2366,& \mathcal{J}^2_{34} = -28.4969. 
\end{eqnarray}
Note that the averages are the same as the desired values.

In the third step we solve for the spanning tree currents, by solving Eqs.(\ref{Eq:SpanTreeCond1},\ref{Eq:SpanTreeCond2}) for $\mathcal{J}^{1,2}_{12}$, $\mathcal{J}^{1,2}_{23}$ and $\mathcal{J}^{1,2}_{24}$. They are given by 
\begin{eqnarray}
\mathcal{J}^1_{12} = -25.4965, & \mathcal{J}^1_{23} =  13.4933,& \mathcal{J}^1_{24} =  9.4902,\\
\mathcal{J}^2_{12} = -21.4965, & \mathcal{J}^2_{23} =  -11.4933,& \mathcal{J}^2_{24} = -7.4902. 
\end{eqnarray} 

In the fourth step we calculate $\mathcal{S}^a$ and $\mathcal{S}^b$ using Eq.(\ref{Eq:equations_for_S}). We present here only the upper part of them as they are symmetric:
\begin{eqnarray}
\mathcal{S}^a = \left(\begin{array}{cccc}
-725.6 & 637.4 & 43& 45.2\\
& -885.2 & 168.7 & 79.1\\
& & -842.1 & 612.4 \\
& & & -736.7
\end{array}\right), \\
\mathcal{S}^b = \left(\begin{array}{cccc}
-596.1 & 474.6 & 89 & 32.5\\
& -682.8 & 143.4 & 64.7\\
& & -1032.6 & 800.2 \\
& & & -897.4
\end{array}\right).
\end{eqnarray} 

In the last step, we first calculate $\partial_t\vec p$ on the two half cycles by proper summation of the currents: $\partial_t\vec p^a = (37.89,-2.51,5.85,-41.22 )$ and $\partial_t\vec p^b = (-37.89,2.51,-5.85,41.22 )$, which as expected cancel each other. These are the slopes of the  $p(t)$'s  in the first and second halves of the period. To keep $0<p(t)<1$, we need to choose $T$ small enough, say for simplicity $T = 0.01$.   Using $T$ we can calculate $\vec p^{a,b}(t)$ which are the actual probability distributions in the two halves. These linear functions are plotted in Fig.(\ref{fig:Solution_vs_Construction_Pt}). Plugging these into Eq.(\ref{Eq:W_final}) gives $\mathcal{W}(t)$.

\section{Zero currents in the NESS}
If $\mathcal{J}_{ij}^{ss}=0$ for some edges, then clearly $\sigma_{ij}^{ss}=0$ for the same edges as well. However, since $\sigma_{ij}^{ps}(t)\geq0$ identically (this follows from Eqs.~(\ref{eq:Jps}) and (\ref{eq:sigmaps})), we must have $\sigma_{ij}^{ps}(t)=0$ for all $t$ in order for the time-averaged entropy production rate of the periodic state to be zero. This can only happen if the currents along these edges are zero at all time, $\mathcal{J}^{ps}_{ij}(t)=0$. A simple prescription to set these currents to zero is to use the  following modified rate matrix: 
\begin{equation} 
 \tilde{\mathcal{R}}_{ij}=(\tilde{\mathcal{S}} + \frac{3}{4}{\mathcal{J}^{ss}})\mathcal{P}^{-1}
\end{equation}
where $\mathcal{P}$ is the matrix with the steady state of $\mathcal{R}$ on its diagonal, $\mathcal{J}^{ss}$ is steady state current matrix of $\mathcal{R}$ (but with different factor in front of it), and 
\begin{eqnarray}
\tilde{\mathcal{S}}_{ij} = \begin{cases}
0 & \mathcal{J}_{ij} = 0 \hbox{ and } i\neq j\\
\frac{3}{2}\mathcal{S}_{ij} & \mathcal{J}_{ij}\neq 0
\end{cases}
\end{eqnarray}
with the diagonal elements of $\tilde{\mathcal{S}}$ changed to make the sum of columns is zero. This modified rate matrix, in which the edges with zero steady state currents have been ``removed'' ($\tilde{\mathcal{R}}_{ij}=0$ on these edges), has the same steady state as $\mathcal{R}$, but its steady state currents are 1.5 larger then those of $\mathcal{R}$. From Eq.(\ref{Eq:sigma_simp}) it also follows that $\tilde{\sigma}^{ss}_{ij} = 1.5 \sigma^{ss}_{ij}$. 

Note that forcing some of the $\tilde{\mathcal{R}}_{ij}$ to be zero might make the time-dependent system  non-ergodic, since it might disconnect some of the states from the others at all times. This, however, can be overcome by dividing the time interval into three equal intervals rather then two. In the first two parts we repeat the construction as before, but using $\tilde{\mathcal{R}}$ instead of $\mathcal{R}$. In the last interval, we choose $\mathcal{W}(t) = \mathcal{S}^c(\Pi^c)^{-1}$, with $\mathcal{S}^c_{ij}=1$ for any $i\neq j$ and $\Pi^c$ the diagonal matrix with $ \vec p^{\,ps}(t=0)$ on its diagonal. With this construction, there are no currents in the periodic solution during the last time interval, the average currents and entropy production rates are the required ones, and the last interval ensures that the system is ergodic.

\end{document}